\documentclass[preprint,12pt]{elsarticle}

\usepackage{lmodern}
\usepackage{amssymb}
\usepackage{url}
\usepackage{amsmath,amssymb,amsfonts}
\usepackage{graphicx}
\usepackage{textcomp}
\usepackage{xcolor}
\usepackage{array}
\usepackage{url}
\usepackage{soul}
\usepackage{color,framed}
\usepackage{color,xcolor}
\usepackage{threeparttable}
\usepackage{multirow}
\usepackage{longtable}
\usepackage{ulem}
\usepackage{algorithm}  
\usepackage{setspace}
\usepackage{array}
\usepackage{booktabs}
\usepackage{rotating}
\usepackage{amsmath}
\usepackage{amssymb}
\usepackage{listings,xcolor}
\usepackage{courier}
\usepackage{caption}   
\usepackage{graphicx}  
\usepackage{caption}
\usepackage{subfigure} 
\usepackage{pdflscape}
\usepackage{bbding}
\usepackage{adjustbox}
\usepackage{enumerate}
\usepackage{blkarray}
\usepackage{comment}
\usepackage{blkarray}
\usepackage{algpseudocode,float}
\usepackage{lipsum}
\usepackage{changepage}
\usepackage{colortbl}
\usepackage{soul}
\usepackage{setspace} 
\newcommand{\tabincell}[2]{\begin{tabular}{@{}#1@{}}#2\end{tabular}}
\newcommand{\reffig}[1]{Fig.\ref{#1}}
\newcommand{\refeqs}[1]{Eq.\ref{#1}}


\begin{document}

\begin{frontmatter}



\title{\textcolor[RGB]{0,0,0}{Minimizing Block Incentive Volatility Through Verkle Tree-Based Dynamic Transaction Storage}}

\author[inst1]{Xiongfei Zhao}
\ead{yb97480@um.edu.mo}

\affiliation[inst1]{organization={University of Macau},
            addressline={Department of Computer and Information Science}, 
            city={Macau}}

\affiliation[inst2]{organization={University of Macau},
            addressline={Center for Data Science}, 
            city={Macau}}
            

\author[inst2]{Gerui Zhang}
\ead{mc15472@um.edu.mo}

\affiliation[inst3]{organization={The Chinese University of Hong Kong},
            addressline={Department of Statistics}, 
            city={Hong Kong}}
            
\author[inst3]{Hou-Wan Long}
\ead{1155190681@link.cuhk.edu.hk}

\fntext[myfootnote]{Corresponding author}
\author[inst1]{Yain-Whar Si\fnref{myfootnote}}
\ead{fstasp@umac.mo}
\begin{abstract}
\fontsize{11.5pt}{23pt}\selectfont  \textcolor[RGB]{0,0,0}{Transaction fees are a crucial revenue source for miners in public and consortium blockchains. However, while public blockchains have additional revenue streams, transaction fees serve as the primary income for miners in consortium blockchains formed by various financial institutions. These miners allocate different levels of computing resources to process transactions and earn corresponding fees. Nonetheless, relying solely on transaction fees can lead to significant volatility and encourage non-standard mining behaviors, thereby posing threats to the blockchain's security and integrity. Despite previous attempts to mitigate the impact of transaction fees on illicit mining behaviors, a comprehensive solution to this vulnerability is yet to be established. To address this gap, we introduce a novel approach that leverages Dynamic Transaction Storage (DTS) strategies to effectively minimize block incentive volatility. Our solution implements a Verkle tree-based storage mechanism to reduce bandwidth consumption. Moreover, to configure the DTS strategies, we evaluate several optimization algorithms and formulate the challenge as a Vehicle Routing Problem. Our experiments conducted using historical transactions from Bitcoin and remittance data from the Industrial and Commercial Bank of China reveal that the strategy focusing on time-based transaction incorporation priority, while excluding a designated space for small-fee transactions, as discovered by the gradient-based optimizer algorithm, proves most effective in reducing volatility. Hence, the DTS strategy can sustain stable block incentives irrespective of transaction types or user bidding behavior. Furthermore, the inclusion of higher-fee transactions, often smaller in size, can alleviate propagation delays and the occurrence of forks.}

\end{abstract}

\begin{keyword}
Blockchain \sep Verkle Tree \sep Transaction Fees \sep Block Incentive \sep Dynamic Transaction Storage \sep Vehicle Routing Problem
\end{keyword}

\end{frontmatter}

\section{Introduction}

\fontsize{11.5pt}{23pt}\selectfont 

\textcolor[RGB]{0,0,0}{Financial institutions prefer to use consortium (i.e., private) blockchains over public ones because consortium blockchains allow participants to preserve the existing profit model while reducing the computational efforts required for block mining activities. Yet, one weakness inherent in traditional consortium blockchain is the absence of mining incentives, built on the assumption that all transaction validators are honest and reliable actors without direct financial motivation. Adequate mining incentives challenge the assumption of rigorous identity verification, pointing to a potential shift towards open and secure blockchain systems \cite{chiu2019blockchain}. However, the integration of system-generated cryptocurrencies into existing financial institutions may pose practical and regulatory challenges.  Our research question specifically addresses whether transaction fees alone can provide sufficient incentives within a consortium blockchain model, thus balancing the need for controlled access with the efficiency of incentive-driven validation.}

\textcolor[RGB]{0,0,0}{Within block mining incentive structures, two key reward mechanisms are recognized: a consistent system-generated cryptocurrencies, and a variable transaction fee reward from users. The first offers miners reliable earnings; the second, our focus, introduces economic complexities due to its fluctuating nature.} According to Carlsten et al. \cite{2978408}, ``with only transaction fees, the block incentive's variance significantly increases due to the exponentially distributed block arrival time, making it appealing to fork a ``wealthy'' block to ``steal'' the rewards contained within.'' Carlsten et al. also identified several deviant mining behaviors such as \textsl{Selfish Mining}, \textsl{Undercutting}, \textsl{Mining Gap}, \textsl{Pool Hopping}, which can disrupt mining stability and harm the ecosystem. \textcolor[RGB]{0,0,0}{Moreover, such deviant mining behaviors impact not only blockchain networks relying solely on transaction fees as their primary mining incentive but also those where transaction fees contribute substantially to the overall block incentives.}

Vitalik Buterin, co-founder of Ethereum, \textcolor[RGB]{0,0,0}{acknowledged via Twitter \cite{vtwitter} that increasing transaction fees on the network might make Ethereum system less secure due to the risks associated with attracting deviant mining behavior. It is also not uncommon to observe spikes in transactions fees in other networks such as Bitcoin. For instance, in May 2023, transaction fees within a single Bitcoin block surpassed the fixed block incentive distributed to miners, totaling 6.7 Bitcoin compared to the block subsidy of 6.25 Bitcoin \cite{TransactionFeesSurpassBlockRewards}.}

\begin{figure}[htp]
    \makeatletter
    \def\@captype{figure}
    \makeatother
    \centering
    \includegraphics[width=0.82 \textwidth]{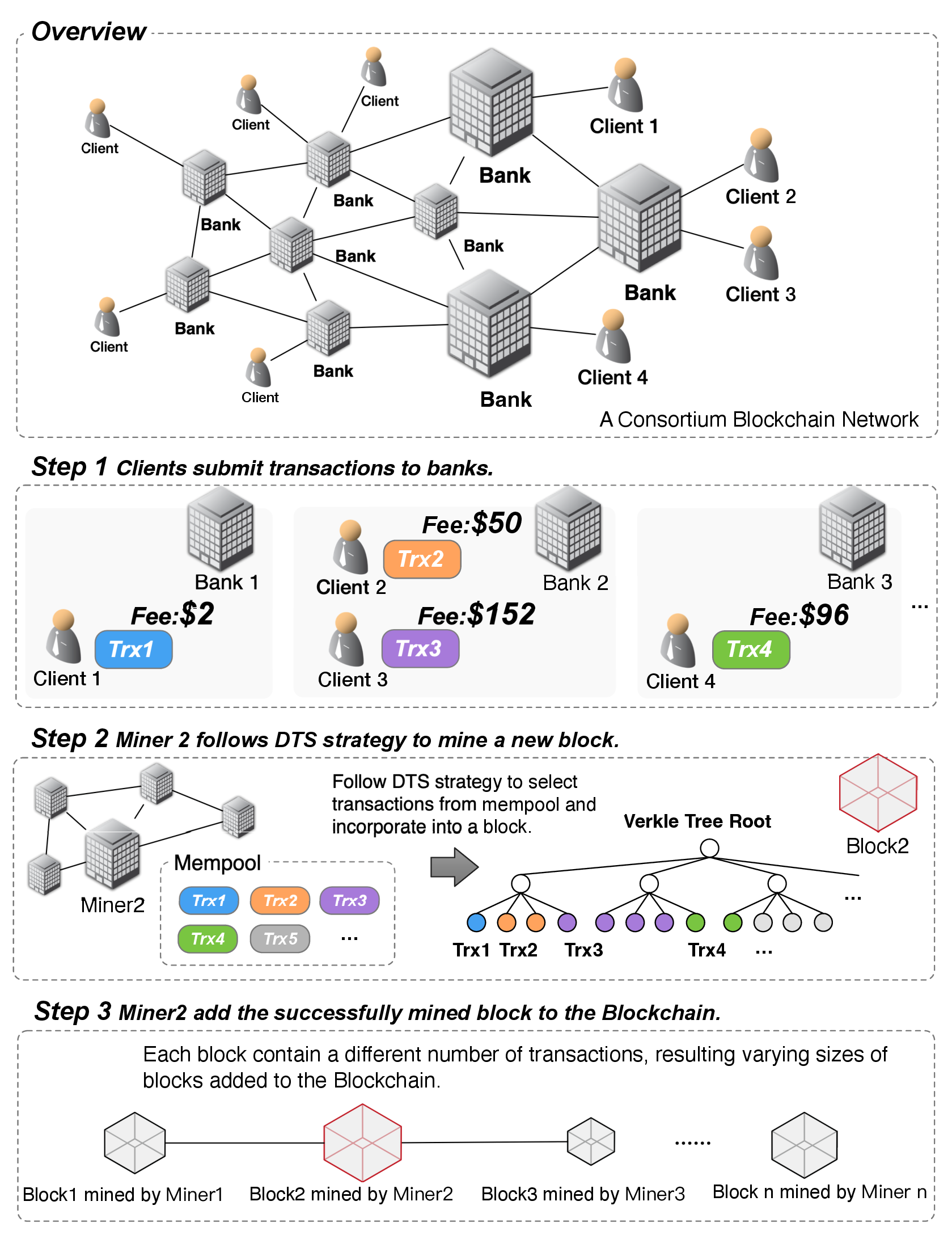}
    \caption{Overview of the proposed approach. The consortium blockchain includes multiple  financial institutions, and user transactions are incorporated using the Dynamic Transaction Storage (DTS) strategy. This approach considers the transaction fee level to determine the Verkle Tree leaf nodes' occupation within a block.}
    \label{VerkleTreeBlockchain} 
\end{figure}

In this study, \textcolor[RGB]{0,0,0}{we introduce a novel consortium blockchain design (illustrated in \reffig{VerkleTreeBlockchain}) to tackle the problem of transaction fee volatility and the potential deviant mining behavior caused by such volatility.} A consortium blockchain is formed through the collaboration of multiple financial institutions. \textcolor[RGB]{0,0,0}{Within the proposed design, the fees acquired by consortium members during transaction settlements serve as crucial incentives for miners contributing to the consortium blockchain. To optimize the use of transaction fees as a mining incentive, our suggested approach is to incorporate a token ecosystem, such as a stablecoin \footnote{Examples of stablecoins that can be used as mining incentives, include US dollar stablecoins such as USDT, Hong Kong dollar stablecoins such as HKDC, and Chinese yuan stablecoins such as CNHC.} that is grounded on a recognized monetary system and enjoys broad acceptance within financial institutions.}

\textcolor[RGB]{0,0,0}{Our proposal is to implement Dynamic Transaction Storage (DTS) strategies to mitigate block incentive volatility and tackle the impact of deviant miner behavior.} The proposed design leverages Verkle Tree data structures to store transactions within blocks, thereby enhancing bandwidth utilization during transaction validation. As depicted in \reffig{VerkleTreeBlockchain}, \textsl{\textbf{Step 1}} involves clients initiating various transactions through different financial institutions, each with varying amounts and transaction fees. \textcolor[RGB]{0,0,0}{In \textsl{\textbf{Step 2}}, for each transaction selected from the Mempool, miners employ the DTS strategy utilizing a Cumulative Distribution Function (CDF).} This function calculates the Verkle Tree leaf nodes occupied by each transaction based on its fee level. Subsequently, once all Verkle Tree leaf nodes are filled, the miner generates a block and disseminates it across the blockchain network, as indicated in \textsl{\textbf{Step 3}}. 

This study presents an innovative solution that incorporates a novel storage mechanism reliant on the Verkle Tree \cite{verkletree}. This addresses the inherent challenges in blockchain implementation, notably reducing robust proof requirements and network bandwidth usage. The Verkle Tree-based mechanism minimizes proof size and bandwidth consumption during transaction verification, thereby augmenting scalability and efficiency, particularly when combined with DTS strategies. \textcolor[RGB]{0,0,0}{Furthermore, we utilize the Vehicle Routing Problem (VRP) framework to abstract the DTS optimization problem.} Employing solution techniques spanning from heuristic algorithms to machine learning technologies within this VRP-based approach facilitates the efficient identification of near-optimal solutions for the DTS strategies.

\textcolor[RGB]{0,0,0}{We employ various advanced optimization algorithms,} including Particle Swarm Optimization (PSO) \cite{eberhart1995new}, differential evolution (DE) \cite{5601760}, Genetic Algorithms (GA) \cite{holland1992genetic}, Covariance Matrix Adaptation Evolution Strategy (CMA-ES) \cite{hansen2001completely}, and Gradient-based Optimizers (GBO) \cite{AHMADIANFAR2020131}, to optimize a DTS strategy. \textcolor[RGB]{0,0,0}{The attributes of the DTS strategy are assessed and determined using these algorithms,} marking a significant advancement in the strategies discussed in previous literature. To validate our approach, extensive experiments were conducted using a combination of these optimization algorithms and a blockchain simulator to identify optimal DTS strategies that minimize volatility. 

In the experiments, we evaluated four categories of DTS strategies aimed at stabilizing volatility. Among these strategies, our findings indicate that the time-based transaction priority strategy, excluding designated space for small-fee transactions, displays the most promising potential for minimizing volatility. 
Our research introduces an innovative approach to designing and operating a consortium blockchain, making several significant contributions to the field:

\textbf{\textsl{1.}}  Our research introduces a novel approach to optimize Verkle tree-based DTS strategies for minimizing block incentive volatility. By redefining the DTS optimization problem within the framework of the VRP, we can utilize established VRP solutions, spanning from heuristic algorithms to machine learning technologies, for effectively identifying near-optimal solutions.

\textbf{\textsl{2.}}  The optimized DTS strategies derived from our study are versatile and can be readily implemented by miners without requiring additional optimization at mining sites. This broad applicability enhances the practical value of our findings and encourages widespread adoption of our strategies across diverse blockchain networks.

\textbf{\textsl{3.}}  The adoption of a Verkle Tree as a transaction storage mechanism in the proposed approach demonstrated a notable decrease in network bandwidth consumption. This enhancement in efficiency aligns with the pressing necessity for optimizing resource utilization in blockchain operations.

The remainder of this paper is structured as follows: Section 2 provides a summary of related work. Section 3 introduces Verkle Tree-Based DTS strategies and outlines the proposed framework for optimizing these strategies. Section 4 details the experimental settings and presents the obtained results. Section 5 presents an analysis on the propagation time and forking rate of the blocks. Finally, Section 6 concludes the paper and discusses avenues for future research.

\section{Related Works}



In \cite{carlsten2016instability}, Carlsten et al. highlighted that ``the incentive issues we uncover arise not because transaction fees may arrive erratically, but because the time-varying nature of transaction fees allows for a richer set of strategic deviations that do not  arise in the block-reward model.'' The term ``time-varying transaction fees'' pertains to fees associated with transactions in the blockchain network that are variable and can fluctuate based on network congestion levels. These fees are influenced not only by the transaction's monetary value but also by market activity during different periods. 


The strategic deviations listed below are directly linked to the volatility of block incentives: The \textbf{\textit{Selfish Mining}} strategy entails miners withholding newly mined blocks until more high-fee transactions accumulate in the Mempool and are included in the block. If block incentive volatility is high, selfish miners can determine the optimal cutoff time. During \textbf{\textit{Undercutting}}, deviant miners actively fork the head of the chain, relinquishing some available unclaimed transaction fees. These unclaimed transactions, associated with fees, can incentivize strategic miners to extend the blocks of deviant miners to garner more rewards.
In \textbf{\textit{Mining Gap}}, amidst high volatility of block incentives, the expected mining reward may fall short of covering mining expenses, rendering it unprofitable for any miner to continue. In \textbf{\textit{Pool Hopping}}, the block incentive is directly linked to the transaction volume in the Mempool. Miners tend to prioritize mining in a Mempool with a higher number of high-fee transactions to maximize block incentives.


Traditional research methodologies in blockchain technology have primarily focused on two levels: the individual transaction level and the overall blockchain level. At the transaction level, studies such as those by Basu et al. \cite{basu2019stablefees} and Matheus et al. \cite{10.1145/3479722.3480991}  have introduced diverse mechanisms aimed at regulating transaction fees. These approaches include methods like the second-price auction-based fee-settling mechanism and a dynamic post-price mechanism. However, these solutions primarily target the fee structure of individual transactions within a block and do not encompass the broader dynamics at the block level.

EIP1559 \cite{EIP1559}, also referred to as the “London Hard Fork,” was activated on Ethereum on August 5, 2021. This update introduces a base fee, representing the minimum gas price necessary for each block. The base fee dynamically adjusts and is burned based on the parent block to alleviate the impact of transaction fees on block incentives. Users can specify transaction fee bids, including a priority fee per gas to incentivize miners for prioritizing their transactions, and a maximum fee per gas to restrict their overall costs. Despite providing a dynamic fee structure, EIP-1559 primarily addresses fees at the transaction level and does not encompass block-level dynamics.



The novel method of this study focuses on the block level, aiming to mitigate systemic blockchain issues like incentive fluctuations. It is the first to apply dynamic storage allocation to reduce block incentive volatility and deter deviant mining behavior.

\section{Verkle Tree-Based Dynamic Transaction Storage Strategies}

Miners' growing dependence on transaction fees might incentivize deviant mining behavior. To ensure the sustainability of the blockchain and prevent deviant mining behaviors, it becomes crucial to stabilize block incentives by flattening the block incentive curve. One potential solution to achieve this stability in block incentives involves dynamically allocating block space based on transaction fees while upholding a consistent block size.

\subsection{\textcolor[RGB]{0,0,0}{DTS}}

Zhao et al. \cite{9592512} introduced multiple novel DTS strategies. Within DTS, the total number of transactions a block can accommodate is dynamically calculated utilizing a CDF based on the transaction fee. The CDF describes the probability of a random variable X being less than or equal to x, representing a probability distribution. In this context, the CDF of the log-normal distribution F(x) is utilized to determine the corresponding transaction fee-based space factor (refer to \refeqs{con:cdf}). Here, $\sigma$ (Shape) denotes the standard deviation of the variable's natural logarithm, and $\mu$ (Scale) represents the expected value or mean.

\begin{equation}
\displaystyle
\resizebox{.5\hsize}{!}{
$F(x) = Pr(X\leq x) = \frac{1}{2} \, + \, \frac{1}{2} \text{erf} \bigg[\frac{ln(x)-\mu}{\sigma\sqrt{2}} \bigg]$ \label{con:cdf}
}
\end{equation}

One of the primary advantages of DTS is its capacity to dynamically regulate the number of transactions that can be included in a block using transaction fees. The calculation of the leaf nodes for a transaction ($T$) involves a formula that considers the transaction fee ($(fee)$) and the maximum leaf nodes designated for a single transaction ($MaxTrxNodes$), as depicted in \refeqs{leaf}. Furthermore, the maximum quantity of leaf nodes allocated for a single transaction can be dynamically adjusted. A larger value indicates a greater number of Merkel Tree leaf nodes reserved for a transaction, while a smaller value implies fewer leaf nodes reserved for a transaction.

{\begin{equation}
\displaystyle
\resizebox{.5\hsize}{!}{
    $LeafNodes(T) = F(fee) \times MaxTrxNodes$ \label{leaf}
}
\end{equation}}


Given that the maximum probability is 1, the maximum value of the CDF is also 1. When the transaction fee reaches a significant level, the upper limit of the maximum number of leaf nodes for a single transaction ($P_{UB}$) approximates the value of $MaxTrxNodes$ (refer to \refeqs{con:upperboundary}). Consequently, the storage space for each transaction can be dynamically allocated, particularly for transactions with extremely high or low transaction fees.

\begin{equation}
    \resizebox{.7\hsize}{!}{
    $P_{UB} = \lim_{fee\to\infty} F(fee) \times MaxTrxNodes = MaxTrxNodes$
    }
    \label{con:upperboundary}
\end{equation}

In formula \refeqs{con:totalincentive}, $I$ is utilized to represent the block's total fee-based incentive. $A$ denotes the individual transaction amount, and $n$ represents the number of transactions incorporated in a block. $f$ signifies the transaction fee, and the block's total incentive $I$ can be calculated using \refeqs{con:totalincentive}. The total number of storage units per block available for incorporating transactions is capped at 2,100.

\begin{equation}
    \resizebox{.23\hsize}{!}{
    $I=\sum_{i=1}^n A_n \times f\%$}
    \label{con:totalincentive} 
\end{equation}

In the Bitcoin Blockchain context, the quantity of transactions permitted in a single block is restricted by the block size. Presently, the maximum block size in Bitcoin is set at 1 MB, equating to roughly 2,100 transactions per block based on the average transaction size. Consequently, the number of transactions $n$ accommodated within a block is limited to 2,100. This restriction is articulated as follows:

\begin{equation}
    \resizebox{.5\hsize}{!}{
    $\sum_{i=1}^n F(A_n \times f\%) \times MaxTrxNodes \leq 2100$
    } 
    \label{con:totalspace} 
\end{equation}

Zhao et al. \cite{9592512} utilized eight attributes in devising DTS strategies, which are briefly outlined below:

\noindent \textbf{\textsl{Mempool Size (A1):}} The Mempool denotes the collection of valid transactions awaiting inclusion in a block. Mempool size refers to the capacity of the holding area where a node retains pending transactions. 

\noindent \textbf{\textsl{Transaction Incorporation Priority (A2):}} This priority is classified into time- or fee-based categories. Time-based transactions follow a first-come, first-served basis for inclusion in blocks. Fee-based transactions prioritize those with higher fees for incorporation into blocks.

\noindent \textbf{\textsl{Designated Space for Small-fee Transactions (A3):}} In earlier Bitcoin phases, 50 KB per block was reserved for high-priority, low-fee transactions. Zhao et al. \cite{9592512} extended this concept by allocating block spaces for transactions with fees falling below specific thresholds. 

\noindent \textbf{\textsl{Small-fee Transaction Fee Threshold (A4):}} This threshold determines a transaction's qualification as a small-fee transaction based on its fee. 

\noindent \textbf{\textsl{Small-fee Transaction Number Threshold (A5):}} This threshold dictates the quantity of small-fee transactions that can be included in a block under DTS strategies. 

\noindent \textbf{\textsl{Maximum Leaf Nodes for One Transaction (A6):}} This attribute regulates the maximum storage space for each transaction, ensuring reasonable allocation even with extraordinarily high transaction fees. 

\begin{figure}[H]
\makeatletter
\def\@captype{figure}
\makeatother
\centering
\includegraphics[width=0.88\textwidth]{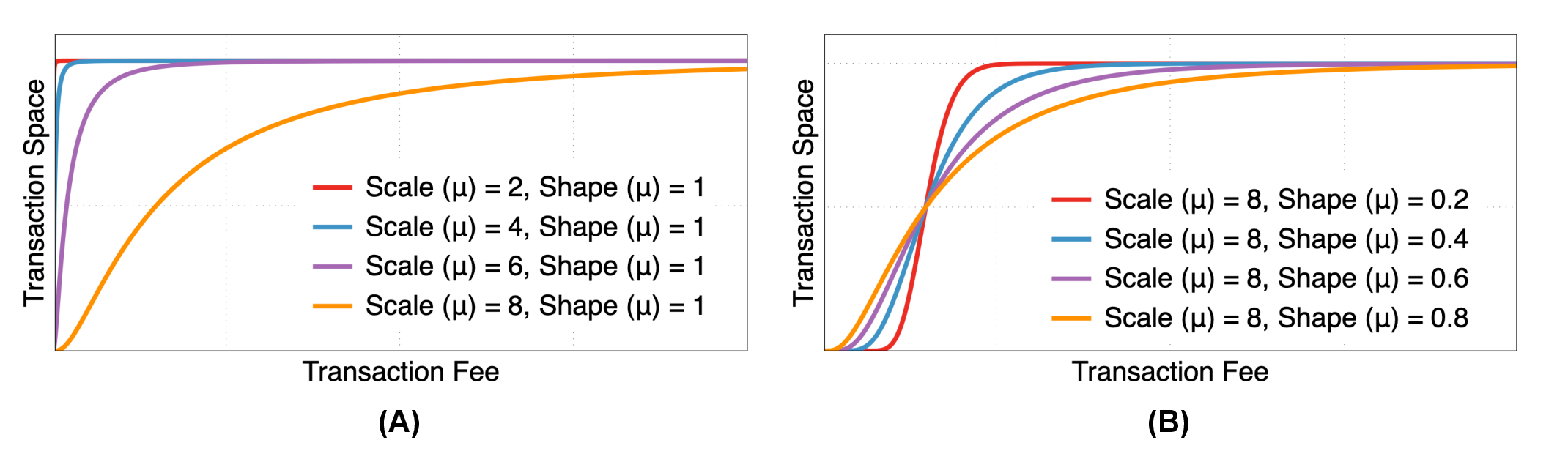}
\caption{Storage space allocation based on transaction fee with different (A) Scale ($\mu$) parameter and (B) Shape ($\sigma$) parameter under DTS strategies}
\label{CDFParam}
\end{figure}

\noindent \textbf{\textsl{Scale (A7):}} In \refeqs{con:cdf}, Scale ($\mu$) represents the mean of the naturally distributed logarithm. This parameter influences the statistical dispersion or ``scale'' of the probability distribution. As depicted in \reffig{CDFParam} (A), a smaller \textsl{Scale} parameter leads to a more concentrated probability distribution, while a larger one results in a more dispersed distribution.

\noindent \textbf{\textsl{Shape (A8):}} The \textsl{Shape} ($\sigma$) in \refeqs{con:cdf} is the standard deviation of the normally distributed natural logarithm. Differing values of the shape parameter under the same transaction fee level lead to distinct probability distributions of the maximum space occupied by transactions, as shown in \reffig{CDFParam} (B).

\begin{figure}[htp]
    \makeatletter
    \def\@captype{figure}
    \makeatother
    \centering
    \includegraphics[width=0.76 \textwidth]{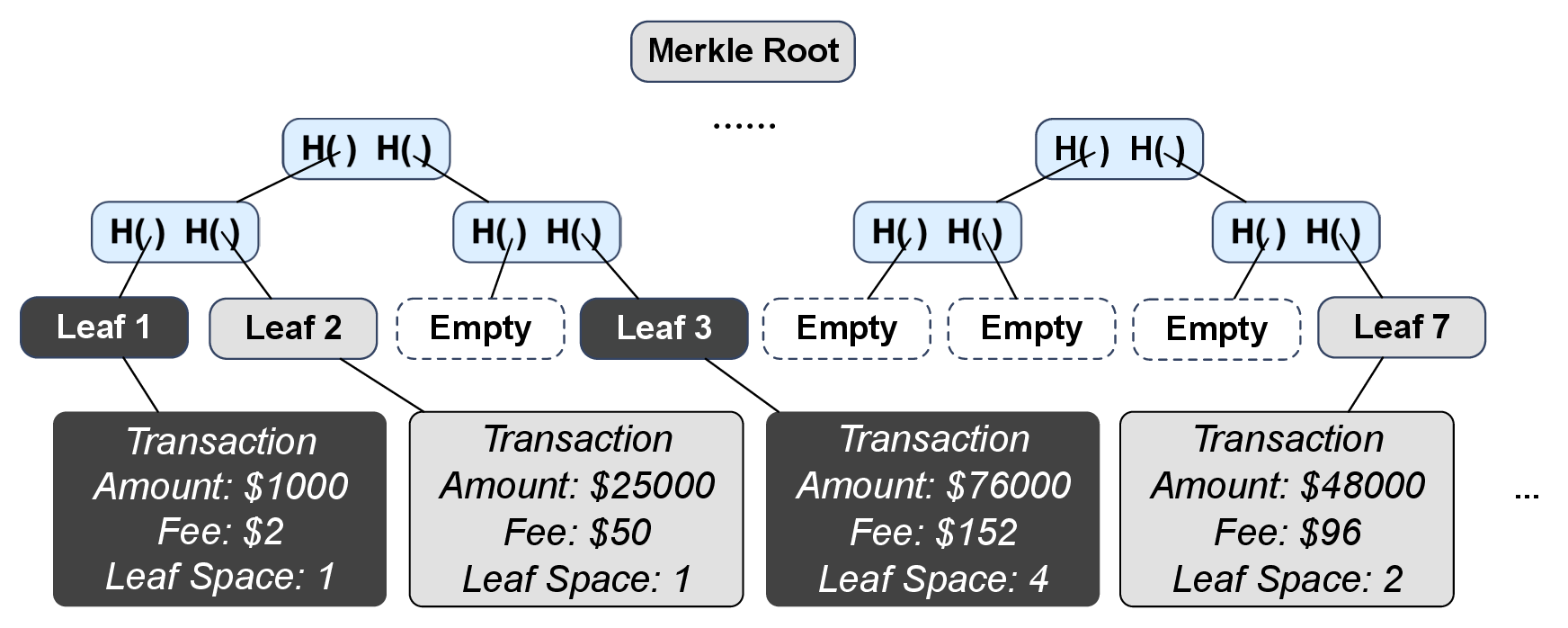}
    \caption{\textcolor[RGB]{0,0,0}{An example of incorporating transactions into a Merkle Tree structure with a branching factor 2 based on DTS strategy \cite{9592512}.}}
    \label{DTS_MerkleTree}
\end{figure}

The simplified concept of the DTS is depicted in \reffig{DTS_MerkleTree}. Initially, the miner calculates the space occupied by each transaction based on its fee using a CDF. Higher transaction fees result in greater utilization of leaf nodes within the Merkel tree structure for the same transaction. 

\subsection{\textcolor[RGB]{0,0,0}{Verkle Tree}}

Currently, the Merkle Tree data structure employs a branching factor of 2, signifying that each node can have up to two child nodes. Verifying whether a transaction belongs to the Merkle Tree involves the Merkle proof, which includes the siblings of each node along the path from the leaf to the Merkle Root (refer to \reffig{DTS_MerkleTree}). Consequently, a Merkle Tree comprising numerous small-leaf nodes may necessitate exceedingly large Merkle proofs. Several existing blockchain scalability solutions, such as XThin \cite{Thinblocks}, Compact \cite{Compactblock}, and Graphene \cite{ozisik2019graphene}, aim to accommodate more transactions within a single block. Table~\ref{ScalabilitySolutions} provides a comparison of proof sizes. In this table, ``Block Capacity'' signifies the number of transactions that can be included in a block, while ``TPS'' represents the transactions per second processed by recent scalability solutions.

As demonstrated in Table~\ref{ScalabilitySolutions}, higher throughput corresponds to larger required proof sizes. Hence, significant network bandwidth is consumed when verifying a substantial number of subsequent transactions. Assuming Bitcoin utilizes a 256-bit hash function like SHA-256 to generate a Merkle Proof, the "Merkle Proof size" (in bytes) can be calculated using \refeqs{con:MerkleTreeDepth}:

\begin{equation}
  \displaystyle
  \resizebox{.3\hsize}{!}{
    $S_m= \frac{ceil(\log_2(N_t)) * 256}{8}$ \label{con:MerkleTreeDepth} 
  }
\end{equation}
where $N_t$ is the number of transactions, and $S_m$ is the ``Merkle Proof size''. 

\begin{table}[htp]
    \footnotesize
  \centering
  \caption{\textcolor{black}{Merkle Proof size comparison of different scalability solutions}}
  \label{ScalabilitySolutions}
\resizebox{1\hsize}{!}{
  \begin{tabular}{p{3.5cm}<{\centering} |p{3cm}<{\centering} |p{3.5cm}<{\centering} |p{4.5cm}<{\centering}}
    \hline\hline
    \specialrule{0em}{2pt}{1pt} 
    \textbf{Scalability Solution} & \textbf{Block Capacity} & \textbf{Throughput (TPS)} & \textbf{Merkle Proof Size (Byte)} \\
    \specialrule{0em}{2pt}{1pt} 
    \midrule
    \specialrule{0em}{2pt}{1pt} 
    Bitcoin & 2100 & 3.5 & \textbf{365.57} \\
    \specialrule{0em}{2pt}{1pt}
    \hline
    \specialrule{0em}{2pt}{1pt} 
    XThin \cite{Thinblocks} & 130,999 & 218.33 & 543.97 \\
    \specialrule{0em}{2pt}{1pt}
    \hline
    \specialrule{0em}{2pt}{1pt}
    Compact \cite{Compactblock} & 174,747 & 291.25 & 557.28 \\
    \specialrule{0em}{2pt}{1pt}
    \hline
    \specialrule{0em}{2pt}{1pt}  
    Graphene \cite{ozisik2019graphene} & 413,507 & 689.17 &  597.04 \\
    \specialrule{0em}{2pt}{1pt}  
    \toprule
  \end{tabular}
  }
\end{table}

\begin{figure}[H]
    \makeatletter
    \def\@captype{figure}
    \makeatother
    \centering
    \includegraphics[width=0.76 \textwidth]{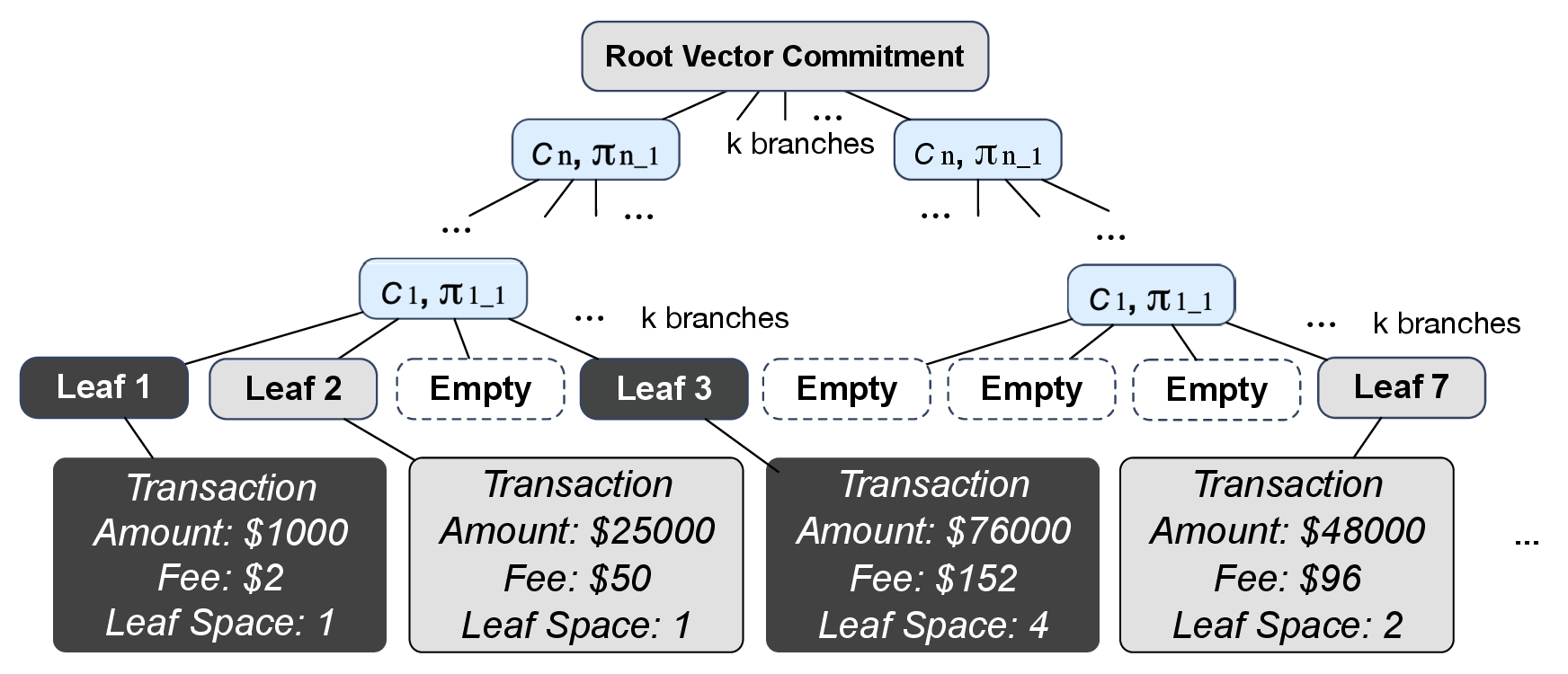}
    \caption{\textcolor[RGB]{0,0,0}{An example of incorporating transactions into a Verkle Tree structure with a branching factor of $k$ based on DTS strategy.}}
    \label{DTS_VerkleTree}
\end{figure}

The Verkle Tree \cite{verkletree}, as proposed by Kuszmaul et al., serves as an alternative to the Merkle Tree, employing a k-ary structure and substituting hash functions with Vector Commitment (VC) \cite{catalano2013vector} to optimize bandwidth utilization. In this structure, the commitment scheme, reliant on the VC cryptographic primitive, operates akin to a digitally sealed envelope, ensuring the security of the selected statement and preserving the original text's integrity. VC permits the commitment to an ordered sequence of values constructed through bilinear groups and the Computational Diffie-Hellman assumption \cite{catalano2013vector}. Illustrated in \reffig{DTS_VerkleTree} is a vertical tree with a branching factor $k$. Transactions are grouped into subsets of size $k$, and a Vector Commitment $C$ is computed for each transaction subset. Following this, membership proofs $\pi$ are calculated for each transaction within the subset concerning $C$. The calculated commitments are utilized iteratively to compute the Vector Commitment until reaching the root Vector Commitment in the tree.

\begin{figure}[H]
  \makeatletter
  \def\@captype{figure}
  \makeatother
  \centering
  \includegraphics[width=0.88\textwidth]{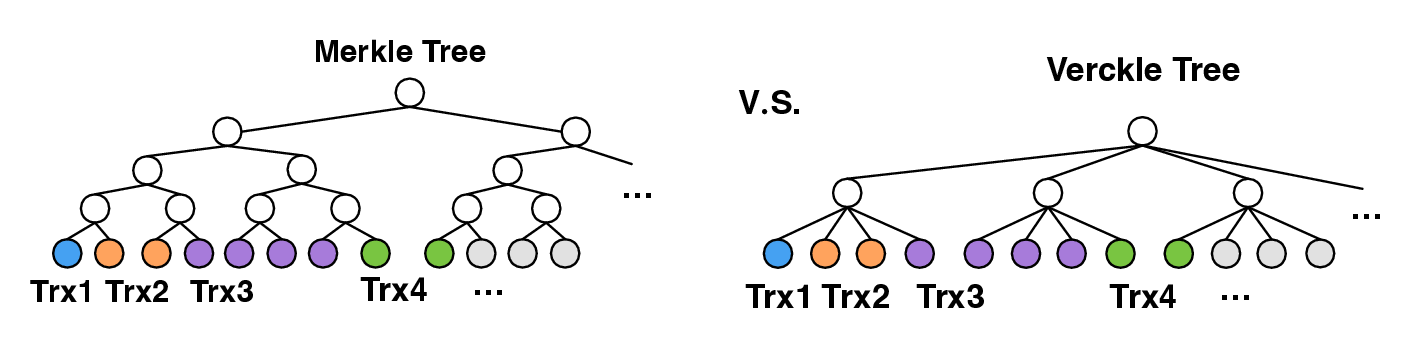}
  \caption{Schematic diagram of the Merkle Tree versus the Verkle Tree with branching factor $k$ = 4}
  \label{VerkleTree}
\end{figure}

Unlike the fixed branching factor of 2 in the Merkle Tree, the Verkle Tree \cite{verkletree} employs a variable branching factor denoted as $k$. Verifying a transaction in a Verkle Tree necessitates an $O(\log_kn)$ membership proof in size. Illustrated in \reffig{VerkleTree}, when the branching factor $k$ equals 2, the Verkle Tree mirrors a binary tree, resembling the structure of the Merkle Tree. However, if both $k$ and the number of transactions $(n)$ are similar, the Verkle Tree transforms into a single-layer tree. Consequently, a larger branching factor $k$ leads to a reduced requirement for membership proof by Vector Commitment for transaction verification within the Verkle Tree.

\textcolor{black}{ By using \refeqs{con:VerkleTreeDepth}, we can calculate the ``Verkle Proof size'' $S_v$ (Byte) based on the number of transactions $N_t$ and branching factor $k$: }

  \begin{equation}
    \resizebox{.32\hsize}{!}{
      $S_v= \frac{ceil(\log_k(N_t)) * 256}{8}$ \label{con:VerkleTreeDepth} 
    }
  \end{equation}

In Table \ref{ScalabilitySolutions2}, we present a comparison of Verkle Proof sizes across different scalability solutions. Establishing a Verkle Tree with a branching factor of $k$ consumes $O(kN_t)$ in terms of computing time \cite{verkletree}. Hence, we selectively include cases where the branching factor is set at 3, 5, and 10, as excessively high branching factors can elevate the computational burden on nodes. Examining Table \ref{ScalabilitySolutions2} reveals that as the branching factor $k$ increases, the required Verkle Proof size decreases. Notably, even for the highest throughput scalability solution (Graphene), a Verkle Proof size of only 257.13 bytes is required when the branching factor $k$ is set to 5. This showcases the Verkle Tree's efficacy in reducing bandwidth consumption compared to the current Bitcoin Merkle Tree (refer to Table \ref{ScalabilitySolutions}). Such bandwidth savings are advantageous when integrating the DTS strategy with solutions like XThin, Compact, Graphene, and other potential high-throughput scalability approaches.

\begin{table}[htp]
    \footnotesize
  \centering
  \caption{\textcolor{black}{Verkle Proof size comparison of different scalability solutions}}
  \label{ScalabilitySolutions2}
\resizebox{1\hsize}{!}{
  \begin{tabular}{p{3.5cm}<{\centering} |p{4cm}<{\centering} |p{4cm}<{\centering} |p{4cm}<{\centering}}
    \hline\hline
    \specialrule{0em}{2pt}{1pt} 
    \textbf{Scalability Solution} & \textbf{Proof Size ($k=3$)} & \textbf{Proof Size ($k=5$)} & \textbf{Proof Size ($k=10$)} \\
    \specialrule{0em}{2pt}{1pt} 
    \midrule
    \specialrule{0em}{2pt}{1pt} 
    Bitcoin & 222.82 & 152.10 & 106.31 \\
    \specialrule{0em}{2pt}{1pt}
    \hline
    \specialrule{0em}{2pt}{1pt} 
    XThin \cite{Thinblocks} & 343.21 & 218.33 & 163.75 \\
    \specialrule{0em}{2pt}{1pt}
    \hline
    \specialrule{0em}{2pt}{1pt}
    Compact \cite{Compactblock} & 351.60 & 240.01 & 167.76 \\
    \specialrule{0em}{2pt}{1pt}
    \hline
    \specialrule{0em}{2pt}{1pt}  
    Graphene \cite{ozisik2019graphene} & 376.69 & \textbf{257.13} & 179.73 \\
    \specialrule{0em}{2pt}{1pt}  
    \toprule
  \end{tabular}
}
\end{table}

In this study, our proposed approach aims to identify an optimal set of Verkle Tree-Based DTS strategies tailored for blockchains. We delineate four primary categories of DTS strategies, which are based on attributes like Transaction Incorporation Priority (A2) and designated space for small-fee transactions (A3). These categories are itemized in Table \ref{DTSStratgies}. Through subsequent experiments, we employ various algorithms to determine the most effective combination of the remaining six attributes (A1, A4, A5, A6, A7, and A8) within each category, intending to minimize block incentive volatility. Moreover, we develop a systematic methodology aimed at mitigating volatility within fee-based block incentives in blockchains, conceptualizing the DTS strategy within the framework of a VRP.

\begin{table}[htp]
    \footnotesize
    \centering
    \caption{\textcolor{black}{DTS Strategy Categories}}
    \label{DTSStratgies}
        \resizebox{0.82\hsize}{!}{
        \begin{tabular}{p{2cm}<{\centering} | p{5cm}<{\centering} | p{5cm}<{\centering}}
        \hline\hline
        \specialrule{0em}{2pt}{1pt}
        \textbf{Strategy Category} & \textbf{Transaction Incorporation Priority (A2)} & \textbf{\textcolor[RGB]{0,0,0}{Designated Space for Small-fee Transactions (A3)}} \\
        \specialrule{0em}{2pt}{1pt}
        \hline
        \specialrule{0em}{2pt}{1pt}
        \tabincell{l}1 & {Time-based} & True \\
        \specialrule{0em}{2pt}{1pt} 
        \hline
        \specialrule{0em}{2pt}{1pt}
        \tabincell{l}2 & {Time-based} & False \\
        \specialrule{0em}{1pt}{1pt} 
        \hline
        \specialrule{0em}{2pt}{1pt}
        \tabincell{l}3 & {Fee-based} & True \\
        \specialrule{0em}{2pt}{1pt} 
        \hline
        \specialrule{0em}{2pt}{1pt}  
        \tabincell{l}4 & {Fee-based} & False \\
        \specialrule{0em}{2pt}{1pt}  
        \toprule
        \end{tabular}
        }
\end{table}

\subsection{Simulator}

SimBlock is a blockchain simulator designed for users to simulate and analyze node behavior's impact on a blockchain. Within the proposed simulation framework, diverse optimization algorithms were assessed to produce a series of candidate solutions representing various combinations of attributes inherent to the DTS strategy. Five optimization algorithms underwent evaluation in this context. They include PSO \cite{eberhart1995new}, DE \cite{5601760}, GA \cite{holland1992genetic}, CMA-ES \cite{hansen2001completely}, and GBO \cite{AHMADIANFAR2020131}. The interaction steps between the optimization algorithm and the blockchain simulator are illustrated in \reffig{PSOonSimBlock}. A summary of these steps is as follows:


\begin{figure}[htp]
  \makeatletter
  \def\@captype{figure}
  \makeatother
  \centering
  \includegraphics[width=0.92\textwidth]{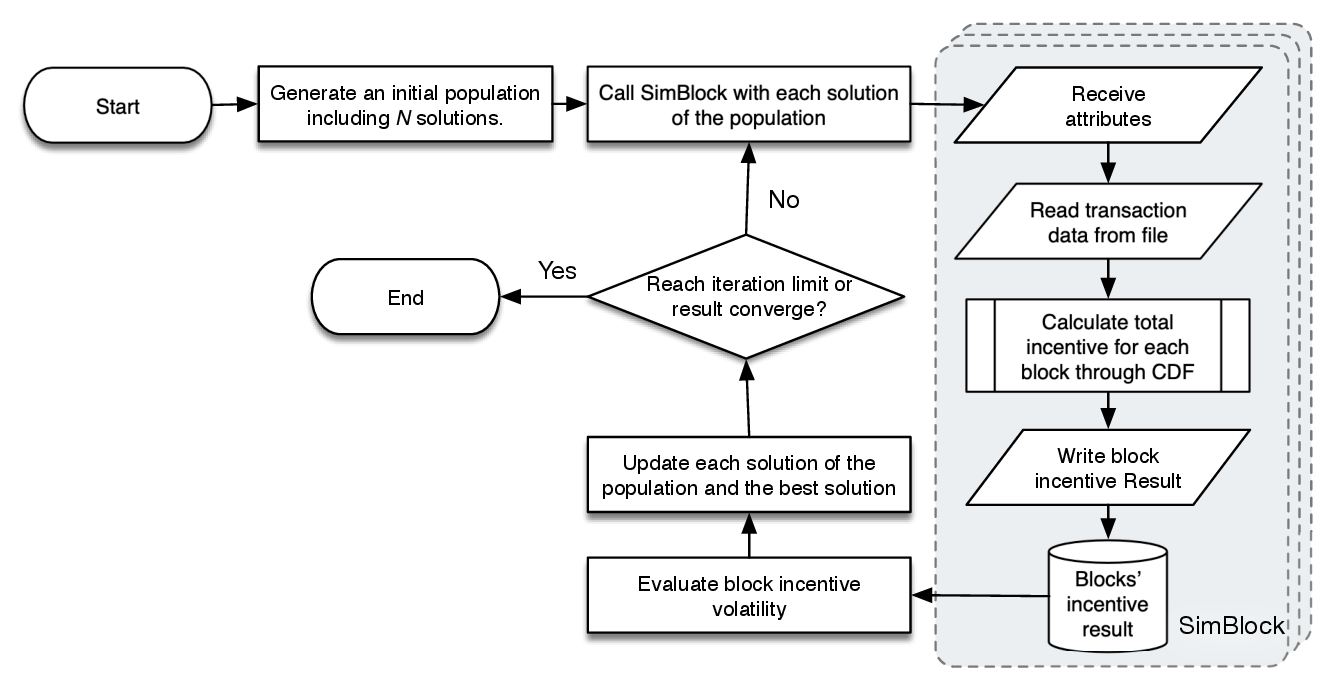}
  \caption{Schematic diagram of the simulation experiment}
  \label{PSOonSimBlock}
\end{figure}

The optimization algorithm commences by initializing candidate solutions, representing diverse attribute combinations for the DTS strategies. These solutions undergo simulation within the simulator to assess their performance. The simulator emulates miner behavior and registers block incentives, considering different DTS attributes. Evaluation of the candidate solutions entails analyzing simulation outcomes and computing the block incentives' volatility. Local best solutions are revised if they exhibit reduced volatility, while the global best solution is correspondingly updated. Optimization iterates by generating new candidate solutions and reiterating the evaluation and update stages until meeting a stopping criterion, such as reaching the maximum number of iterations or achieving the desired convergence.

\subsection{Vehicle Routing Problem (VRP)}

The VRP stands as a widely recognized optimization challenge within the information systems domain. This problem constitutes a combinatorial optimization task involving the identification of the most efficient routes for a fleet of vehicles to service a designated set of customers. The DTS strategy can be likened to a VRP due to their shared elements of optimization and resource allocation. Within the DTS framework, transactions resemble customers in VRP, each presenting specific requirements—such as incentives and storage in DTS, or goods to be transported in VRP. In DTS, a block plays a role akin to a vehicle in VRP, responsible for incorporating transactions within its storage capacity, just as a vehicle must deliver goods within its carrying limit. Both scenarios share a common optimization objective—to minimize block incentive volatility in DTS and to minimize overall travel time or distance in VRP.

Further parallels exist between the DTS and VRP. The transaction storage mechanism within DTS, determining the storage allocation for each transaction within a block, can be likened to the various vehicle capacities seen in the VRP. Similarly, the transaction incorporation priority within DTS, influencing the sequence of transaction inclusion, bears resemblance to the prioritization of deliveries within the VRP. This abstraction enables the application of established VRP solution methodologies to address the DTS issue, encompassing heuristic algorithms and machine learning techniques, enabling effective and efficient determination of near-optimal solutions. In the subsequent sections, we outline the formulation of DTS strategies within the context of a VRP.

\textbf{\textsl{Customers (transactions)}}: Customers (transactions): In our approach, each transaction within the network is considered akin to a customer in the VRP, with the demand of each customer corresponding to the transaction size. The geographical positioning of each customer was determined based on their fees and arrival times.

\begin{figure}[H]
\makeatletter
\def\@captype{figure}
\makeatother
\centering
\includegraphics[width=0.82\textwidth]{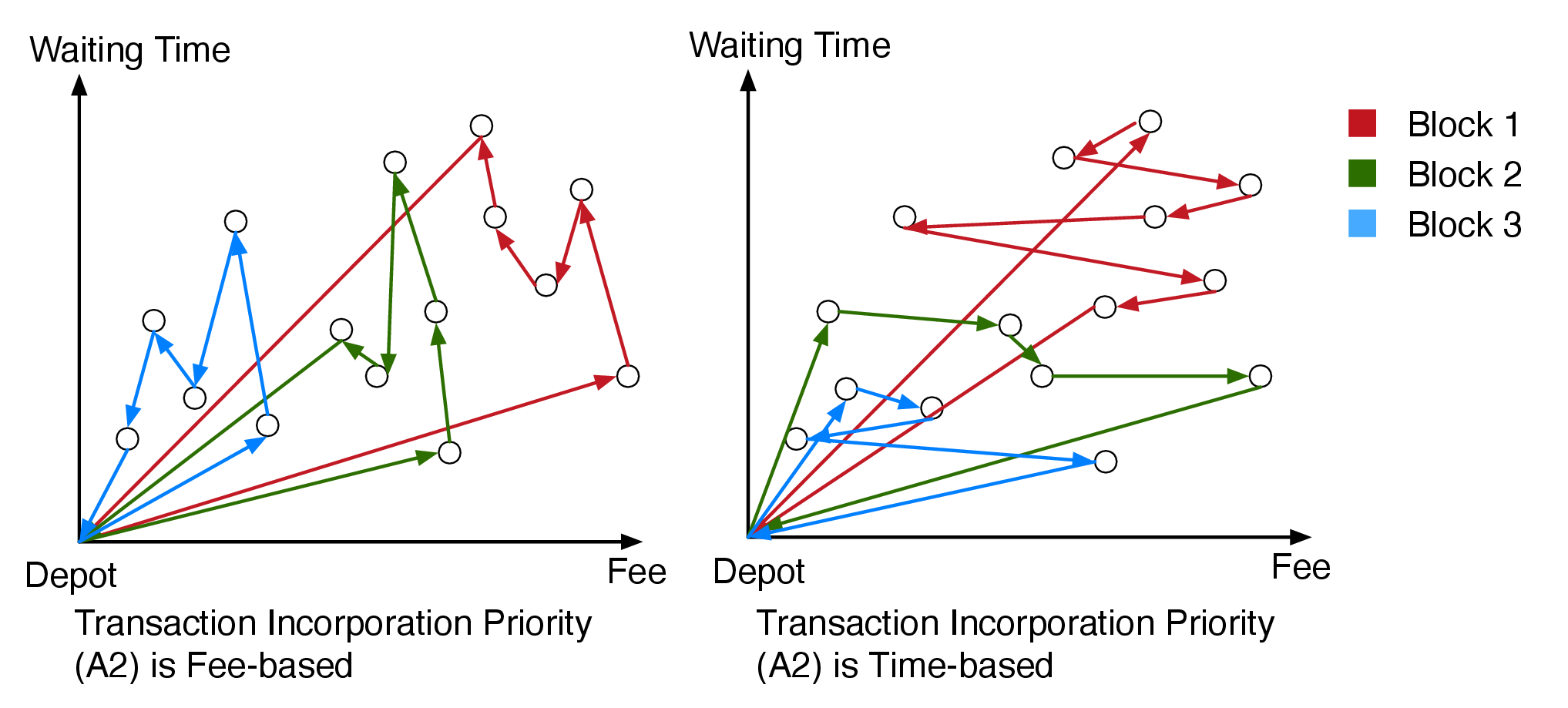}
\caption{Time-based selection versus Fee-based selection in TRP}
\label{DTS_TRP}
\end{figure}

The classification of DTS strategies is grouped into four primary categories by considering two fundamental attributes: Transaction Incorporation Priority (A2) and Designated Space for Small-fee Transactions (A3), as outlined in Table \ref{DTSStratgies}. Within the context of the VRP, we explore various values of parameter A2, while maintaining A3 at a constant value of \textsl{False}. This scenario is exemplified in rows 2 and 4 of Table \ref{DTSStratgies}. When the A2 priority attribute is designated as ``Fee-based,'' the order of customer selection in the VRP is contingent upon their transaction fees. Essentially, customers offering higher transaction fees are given priority during the selection process (depicted in the left subfigure of \reffig{DTS_TRP}). Conversely, when the A2 priority attribute is set to ``Time-based,'' selection priority is based on the waiting duration of customers (illustrated in the right sub-figure of \reffig{DTS_TRP}). This implies that customers with longer waiting times receive priority during the selection process.


\textbf{\textsl{Vehicles (blocks)}}: The blocks within the blockchain network are analogous to vehicles in the VRP, with the capacity of each vehicle representing its maximum block size. This capacity can dynamically change depending on the parameters of the aforementioned DTS strategy. \textbf{\textsl{Cost (transaction fee)}}: The cost element is not defined by the geographical distance between two customers, but rather by the transaction fee required for a customer to be included in a block. Unlike traditional VRP, which aims to minimize the total distance traveled, our objective is to minimize the volatility of block incentives. \textbf{\textsl{Depot}}: In the DTS strategy, the depot fulfills a role akin to the starting location in a traditional VRP. Specifically, it serves as the initial point for each block's journey, commencing with block mining. The depot in this context is characterized by a fixed block incentive level that must be reached to ensure successful mining of the block, thereby minimizing the volatility of the block incentives.

\textcolor[RGB]{0,0,0}{We then formulate the modified VRP as follows:}

\begin{equation}
\label{trx_selection}
    \resizebox{.9\hsize}{!}{\textcolor[RGB]{0,0,0}{$I(A1,A2,A3,A4,A5,A6,A7,A8)=\{x_{i, k}\},\quad x_{i, k}\in \{0,1\},\quad\forall i\in Trx,\quad k\in Block$}}
\end{equation}

\begin{equation}
    \label{vrp1}
    \resizebox{.5\hsize}{!}{\textcolor[RGB]{0,0,0}{$min(Var\{\sum_{i\in Trx}(c_i*x_{i, k})\quad|\quad k\in Block\quad\})$}}
\end{equation}

\begin{equation}
    \label{constrain1}
    \resizebox{.35\hsize}{!}{\textcolor[RGB]{0,0,0}{$\sum_{i\in Trx}x_{i, k}=1,\quad \forall k\in Block$}}
\end{equation}

\begin{equation}
     \label{constrain2}
    \resizebox{.35\hsize}{!}{\textcolor[RGB]{0,0,0}{$\sum_{k\in Block}x_{i, k}=1,\quad \forall i\in Trx$}}
\end{equation}

In our formulations, $c_{i}$ represents the fee associated with packing the \textit{i-th} transaction into a block, while $x_{i, k}$ is a binary variable taking the value of one if block $k$ contains transaction $i$. Equation 9 allocates transactions to their corresponding blocks based on a specified DTS strategy. Equation 10 represents the objective of minimizing the variance (denoted as $Var$) of the set of block incentives. \refeqs{constrain1} and \refeqs{constrain2} impose constraints to ensure that each transaction is packed and can only be packed once within a given block. These formulas establish a rigorous mathematical foundation for the modified VRP approach.

\section{Experiments}

\subsection{Experiment Settings}

SimBlock \cite{8751431} was employed to assess the effectiveness of DTS strategies in minimizing block incentive volatility. In our experiment, a commission ratio of 0.2\% was assumed for each user-initiated transaction. The parameter settings of the SimBlock Simulator encompass various fundamental variables. These include the block size, currently set at 1 MB; the number of transactions permitted in one block, established at 2,100; and the rate at which new transactions are added to the Mempool, set at 3.5 transactions per second.


The parameters of the five optimization algorithms are listed in Table \ref{opt_param}. All optimization algorithms, except CMA-ES, utilized a population of 50 and a maximum generation of 100. CMA-ES had a maximum number of solutions to be evaluated set at 5000, equivalent to the product of the population and maximum generation, ensuring optimization efficiency and effectiveness. Clerc and Kennedy \cite{clerc2002particle} proposed deriving PSO parameters from \refeqs{pso_param}, wherein $\chi$  represents the constriction coefficient. Following \cite{clerc2002particle}'s recommendations, we set $k$ = 1, $\phi_{1} = \phi_{2}$ = 2.05 to calculate $w$, $c_1$ and $c_2$. For DE, the crossover constant was set to 0.9, a more suitable value depending on the attributes to be tuned. Regarding GBO, its local escape probability was established at 0.5, following guidance from \cite{AHMADIANFAR2020131}.

\begin{equation}
    \resizebox{.26\hsize}{!}{$\left\{
    \begin{aligned}
    w = & \chi \\\
    c_1 = & \phi_1 \times \chi \\\
    c_2 = & \phi_2 \times \chi \\\
    \chi = & \frac{2k}{|2-\phi-\sqrt{\phi^2-4\phi}|}
    \end{aligned}
    \right.$}
\label{pso_param}
\end{equation}

\textcolor[RGB]{0,0,0}{where $0\leq k\leq 1$ and $\phi = \phi_{1}+\phi_{2}\geq 4$.}

\begin{table}[htp]
\footnotesize
\caption{\textcolor{black}{Parameters for Optimization Algorithms}}
\label{opt_param}
\centering
\resizebox{0.92\hsize}{!}{
\begin{tabular}{>{\color[RGB]{0,0,0}}p{1.8cm}<{\centering} | >{\color[RGB]{0,0,0}}p{5cm}<{\centering} | >{\color[RGB]{0,0,0}}p{1cm}<{\centering} | >{\color[RGB]{0,0,0}}p{5cm}<{\centering} | >{\color[RGB]{0,0,0}}p{1cm}<{\centering}}
\hline\hline\textbf{Algorithm} & \textbf{Hyper-parameter} & \textbf{value} & \textbf{Hyper-parameter} & \textbf{value} \\ 
\specialrule{0em}{0pt}{0pt}
\hline

\multirow{5}{*}{PSO}                  
  & population ($n\_pop$) & 50 & maximum generation ($max\_gen$) & 100 \\ 
  \specialrule{0em}{2pt}{1pt}\cline{2-5}
  & inertia weight ($w$) & 0.73 & acceleration coefficient ($c1$) & 1.50  \\ 
  \specialrule{0em}{2pt}{1pt}\cline{2-5}
  & acceleration coefficient ($c2$) & 1.50 &   &  \\ 
  \specialrule{0em}{2pt}{1pt}\hline
\multirow{3}{*}{DE}  
  & population ($n\_pop$) & 50 & maximum generation ($max\_gen$) & 100 \\ 
  \specialrule{0em}{2pt}{1pt}\cline{2-5}
  & crossover constant ($CR$) & 0.9 &    &  \\ 
  \specialrule{0em}{2pt}{1pt}\hline
\multirow{1}{*}{GA}  
  & population ($n\_pop$) & 50 & maximum generation ($max\_gen$) & 100 \\ 
  \specialrule{0em}{2pt}{1pt}\hline
CMA-ES               
  & number of evaluations ($n\_eval$) & 5000 &   &  \\ 
  \specialrule{0em}{2pt}{1pt}\hline
\multirow{3}{*}{GBO} 
  & population ($n\_pop$) & 50 & maximum generation ($max\_gen$) & 100  \\ 
  \specialrule{0em}{2pt}{1pt}\cline{2-5}
  & local escaping probability ($prob$) & 0.5 &    &  \\ 
  \specialrule{0em}{2pt}{1pt}\hline
\end{tabular}
}
\end{table}

\subsection{Benchmarks}

The choice of volatility as a measure to assess the effectiveness of our proposed strategy stems from its common usage in describing the degree of variation or uncertainty within mining incentives. Elevated volatility within block incentives can result in sharp or frequent fluctuations, both of which heighten the potential for deviant mining behaviors. Such fluctuations may prompt individual miners to engage in deviant practices. Hence, reducing the volatility of block incentives can foster a more stable and secure mining ecosystem.

The standard deviation of logarithmic returns is commonly employed as a measure of volatility \cite{doi:10.2469/faj.v46.n3.23}. In this study, we adopted a similar methodology to gauge fluctuations in block incentives. Assuming the total block incentive is represented by $I_n$, we derived $R_{n}$ using the following equation:

\begin{equation}
    \resizebox{.2\hsize}{!}{$R_{n}=ln(I_{n}/I_{n-1})$}\label{reward}
\end{equation}

To compute the standard deviation of block incentives, we determined the average $R_{n}$. The average continuously compounded block incentive, denoted as $R_{avg}$ , is defined as follows:

\begin{equation}
    \resizebox{.2\hsize}{!}{$R_{avg}=\frac{\sum_{i=1}^nR_{i}}{n}$}\label{average}
\end{equation}

where $\sigma$ is the standard deviation of block incentives in a particular period. By employing a rolling window, which encompasses a period of $n$ consecutive blocks culminating at the last block, we calculated $\sigma$ in the following manner:

\begin{equation}
    \resizebox{.3\hsize}{!}{$\sigma = \sqrt{\frac{\sum_{i=1}^n(R_{i}-R_{avg})^{2}}{n-1}}$}\label{standard deviation}
\end{equation}

We obtained daily data on total cryptocurrency-based block incentives and transaction fees paid to miners from the NASDAQ website, covering a continuous period of 365 blocks for each year spanning from 2012 to 2021. Subsequently, we defined the block incentive from these records as $R$ and downloaded the number of blocks $B$ mined on the Bitcoin network daily from 2012 to 2020 using BTC.com. Employing the formula $A_{n}=R_{n}/B_{n}$, we computed the average block incentive for each day, denoted as $A_{n}$ (where $n=365$ for each year). Next, we calculated the day-to-day returns $R$ using the formula $R_{n}=ln(A_{n}/A_{n-1})$. Finally, we determined the standard deviation of the returns $R_{n}$ using the following formula: $ \sigma = \sqrt{\frac{\sum_{i=1}^n(R_{i}-R_{avg})^{2}}{n-1}}$, where $R_{avg}$ is the average of $R_{n}$ defined as $\frac{\sum_{i=1}^nR_{i}}{n}$. 

The historical volatility values of Bitcoin’s average daily block incentives from 2012 to 2021 are presented in Table \ref{HistoricalVolatility}. The block incentive's volatility ranged from 0.037647 in 2019 to 0.238111 in 2012, and the minimum and maximum values were utilized as the reference volatility ranges in our experiments.

\begin{table}[htp]
  \centering
  \caption{\textcolor{black}{Block incentives' historical volatility from 2012 to 2021}}
  \resizebox{0.9\hsize}{!}{
  \begin{threeparttable}
    \begin{tabular}{>{\color[RGB]{0,0,0}}p{1cm}<{\centering}|>{\color[RGB]{0,0,0}}p{4.2cm}<{\centering}|>{\color[RGB]{0,0,0}}p{1cm}<{\centering}|>{\color[RGB]{0,0,0}}p{4.2cm}<{\centering}|>{\color[RGB]{0,0,0}}p{1cm}<{\centering}|>{\color[RGB]{0,0,0}}p{4.2cm}<{\centering}}
      \hline\hline
      \textbf{Year} & \textbf{Historical Volatility} & \textbf{Year} & \textbf{Historical Volatility} & \textbf{Year} & \textbf{Historical Volatility}\\
      \specialrule{0.00em}{2pt}{1pt} 
      \hline
      2020 & 0.059485 & 2017 & 0.063965 &  2014 & 0.218010 \\
      \specialrule{0.00em}{2pt}{1pt} 
      \hline
      \textbf{2019} & \textbf{0.037647} & 2016 & 0.073051 &  2013 & 0.200857\\
      \specialrule{0.00em}{2pt}{1pt} 
      \hline
      2018 & 0.045616 & 2015 & 0.180948 & \textbf{2012} & \textbf{0.238111}\\
      \specialrule{0.00em}{2pt}{1pt}   
      \toprule
    \end{tabular}
  \end{threeparttable}
  }
  \label{HistoricalVolatility}
\end{table}

\subsection{\textcolor{black}{Experimental Results}}


Table \ref{optimization_res} presents the results obtained from five different optimization algorithms, each having four experiments conducted. The findings highlight that Experiment 17, specifically the “Time-based transaction incorporation priority without a designated space for small-fee transactions,” implemented via the GBO optimization algorithm, emerges as the most promising approach for stabilizing block incentives. This determination is based on its notably lower volatility level of 0.1158, which stands as the smallest among all the experiments conducted.

\begin{table}[H]
\footnotesize
\caption{\textcolor[RGB]{0,0,0}{Optimization Results with PSO, DE, GA, CMA-ES, and GBO}}
\label{optimization_res}
\centering
\resizebox{0.96\hsize}{!}{
\begin{threeparttable}
\begin{tabular}{>{\color[RGB]{0,0,0}}p{1.4cm}<{\centering} >{\color[RGB]{0,0,0}}p{2.6cm}<{\centering} >{\color[RGB]{0,0,0}}p{0.8cm}<{\centering} >{\color[RGB]{0,0,0}}p{2.2cm}<{\centering} >{\color[RGB]{0,0,0}}p{0.6cm}<{\centering} >{\color[RGB]{0,0,0}}p{0.6cm}<{\centering} >{\color[RGB]{0,0,0}}p{0.6cm}<{\centering} >{\color[RGB]{0,0,0}}p{0.6cm}<{\centering} >{\color[RGB]{0,0,0}}p{0.6cm}<{\centering} >{\color[RGB]{0,0,0}}p{0.6cm}<{\centering} >{\color[RGB]{0,0,0}}p{1.4cm}<{\centering}}
\hline\hline
\textbf{Algorithm} & \textbf{Experiment} & \textbf{A1} & \textbf{A2} & \textbf{A3} & \textbf{A4} & \textbf{A5} & \textbf{A6} & \textbf{A7} & \textbf{A8} & \textbf{Volatility} \\ 
                     \specialrule{0em}{2pt}{0pt}\hline
\multirow{4}{*}{PSO} & Experiment 1  & 32569 & Time-based & False & -    & -   & 77   & 6.43 & 0.73  & 0.1338     \\ 
                     \specialrule{0em}{2pt}{0pt}\cline{2-11} 
                     & Experiment 2  & 27443 & Time-based & True  & 1.41 & 110 & 93   & 6.72 & 0.91  & 0.1192     \\ 
                     \specialrule{0em}{2pt}{0pt}\cline{2-11} 
                     & Experiment 3  & 48758 & Fee-based  & False & -    & -   & 65   & 6.49 & 0.91  & 0.3139     \\ 
                     \specialrule{0em}{2pt}{0pt}\cline{2-11} 
                     & Experiment 4  & 72146 & Fee-based  & True  & 1.27 & 161 & 10   & 5.32 & 0.16  & 0.4285     \\ 
                     \specialrule{0em}{2pt}{0pt}\hline
\multirow{4}{*}{DE}  & Experiment 5  & 27094 & Time-based & False & -    & -   & 90   & 6.75 & 0.97  & 0.1192     \\ 
                     \specialrule{0em}{2pt}{0pt}\cline{2-11}
                     & Experiment 6  & 6317  & Time-based & True  & 1.21 & 89  & 103  & 6.89 & 1.00  & 0.1165     \\ 
                     \specialrule{0em}{2pt}{0pt}\cline{2-11}
                     & Experiment 7  & 68311 & Fee-based  & False & -    & -   & 82   & 6.81 & 0.99  & 0.2902     \\ 
                     \specialrule{0em}{2pt}{0pt}\cline{2-11}
                     & Experiment 8  & 75004 & Fee-based  & True  & 1.12 & 177 & 31   & 4.70 & 0.98  & 0.2032     \\ 
                     \specialrule{0em}{2pt}{0pt}\hline
\multirow{4}{*}{GA}  & Experiment 9  & 70714 & Time-based & False & -    & -   & 127  & 7.01 & 1.00  & 0.1167     \\ 
                     \specialrule{0em}{2pt}{0pt}\cline{2-11}
                     & Experiment 10 & 45492 & Time-based & True  & 1.06 & 36  & 112  & 6.88 & 0.97  & 0.1194     \\ 
                     \specialrule{0em}{2pt}{0pt}\cline{2-11}
                     & Experiment 11 & 73618 & Fee-based  & False & -    & -   & 62   & 6.59 & 0.99  & 0.3069     \\ 
                     \specialrule{0em}{2pt}{0pt}\cline{2-11}
                     & Experiment 12 & 74535 & Fee-based  & True  & 1.02 & 97  & 37   & 5.16 & 1.00  & 0.2106     \\ 
                     \specialrule{0em}{2pt}{0pt}\hline
\multirow{4}{*}{CMA-ES} & Experiment 13 & 49325 & Time-based & False & -    & -  & 122 & 7.02 & 1.00 & 0.1162     \\ 
                     \specialrule{0em}{2pt}{0pt}\cline{2-11} 
                     & Experiment 14 & 62662 & Time-based & True  & 1.02 & 114 & 110  & 6.93 & 0.99  & 0.1183     \\ 
                     \specialrule{0em}{2pt}{0pt}\cline{2-11} 
                     & Experiment 15 & 74972 & Fee-based  & False & -    & -   & 759  & 9.70 & 0.47  & 0.4855     \\ 
                     \specialrule{0em}{2pt}{0pt}\cline{2-11}
                     & Experiment 16 & 1141  & Fee-based  & True  & 1.05 & 3   & 34   & 7.26 & 0.26  & 0.5186     \\ 
                     \specialrule{0em}{2pt}{0pt}\hline
\multirow{4}{*}{GBO} & \textbf{Experiment 17} & \textbf{25469} & \textbf{Time-based} & \textbf{False} & \textbf{-}    & \textbf{-}   & \textbf{110}  & \textbf{6.94} & \textbf{1.00}  & \textbf{0.1158}     \\ 
                     \specialrule{0em}{2pt}{0pt}\cline{2-11}
                     & Experiment 18 & 75039 & Time-based & True  & 1.32 & 1   & 110  & 6.92 & 1.00  & 0.1178     \\ 
                     \specialrule{0em}{2pt}{0pt}\cline{2-11}
                     & Experiment 19 & 73714 & Fee-based  & False & -    & -   & 79   & 6.79 & 0.99  & 0.2920     \\ 
                     \specialrule{0em}{1pt}{0pt}\cline{2-11}
                     & Experiment 20 & 71907 & Fee-based & True & 1.47 & 17 & 56   & 6.19 & 1.00  & 0.2595   \\
                     \specialrule{0em}{2pt}{0pt}
\toprule
\end{tabular}
\begin{tablenotes}
\footnotesize
\item A1: Mempool Size; A2: Transaction Incorporation Priority; A3: Designated Space for Small-fee Transactions; A4: Small-fee Transaction Fee Threshold; A5: Small-fee Transaction Number Threshold; A6: Maximum Leaf Nodes for One Transaction; A7: Scale; A8: Shape;
\end{tablenotes}
\end{threeparttable}
}

\end{table}

\subsection{\textcolor[RGB]{0,0,0}{Analysis on Proof Size}}

To meet the transaction throughput demands in financial applications, the optimized DTS strategy necessitates complementary integration with Segwit \cite{lombrozo2015segregated}, Compact Blocks \cite{Compactblock}, Graphene \cite{ozisik2019graphene}, and other scalable solutions. However, this integration leads to an accumulation of numerous transactions in a block, requiring a substantial number of Merkle proofs for transaction validation. Suppose the DTS strategy is integrated with a Graphene solution to enhance scalability. This integration allows a block to accommodate a maximum of 540,000 transactions, resulting in a Merkle Tree with a depth of $\log_{2}{540,000}$ levels, approximately 18.1. Hence, the tree would consist of 18 levels of non-leaf nodes and one additional level for the root node, totaling 19 levels. Each Merkle proof comprises one node, a 256-bit hash such as SHA-256, at every tree level. Consequently, the size of the Merkle proof for a block is calculated as $19 * 32 \text{ bytes} = 608 \text{ bytes}$, double the size of a typical bitcoin transaction (approximately 300 bytes). In contrast, Verkle Trees decrease the proof size from $O(\log_{2}{n})$ to $O(\log_{k}{n})$ for a branching factor of $k$. This reduction results in $\log_{2}{k}$ less bandwidth usage. Thus, if we set $k = 1024$, the Verkle-proof size shrinks tenfold, measuring only 60 bytes, significantly smaller than the 608 bytes required for a Merkle proof.

\subsection{\textcolor[RGB]{0,0,0}{Experiments for DTS Strategy with Additional Dataset}}

The experiment is based on 400,000 remittances spanning January 2022 to December 2022, gathered from the ICBC Macau, along with 400,000 historical market data entries covering Bitcoin from December 2019 to September 2020. 
We employed the most effective attribute settings of the DTS strategy, namely, ``Time-based transaction incorporation priority without a designated space for small-fee transactions'' (refer to Table \ref{optimization_res}, Experiment 17). These attribute settings were utilized in SimBlock to simulate the incorporation of all transactions into blocks. Subsequently, we derived the volatility of fee-based incentives for different blocks. The data in Table \ref{IrrUsers} distinguishes between two scenarios: Rows 1 and 3 represent the baseline of transactions in the absence of irrational users, while rows 2 and 4 depict the incorporation of transactions involving irrational users. The final column of Table \ref{IrrUsers} illustrates the volatility of the block incentives corresponding to varying ratios of rational and irrational users. The results from the experiments in Table \ref{IrrUsers} (rows 1 and 3) indicate that, with rational miners, the DTS strategy can maintain the volatility of block incentives within the historical range obtained from Table \ref{HistoricalVolatility} (i.e., 0.037647–0.238111). Furthermore, observations from Table \ref{IrrUsers} (rows 2 and 4) for scenarios involving irrational miners suggest that the DTS strategy can also sustain the volatility of block incentives within the historical range (i.e., 0.037647 to 0.238111).

\begin{table}[htb]
\footnotesize
\caption{\textcolor{black}{Experiment Results for DTS Strategy with Additional Dataset}}
\label{IrrUsers}
\centering
\resizebox{0.96\hsize}{!}{
\begin{threeparttable}
\begin{tabular}{>{\color[RGB]{0,0,0}}p{0.5cm}<{\centering} | >{\color[RGB]{0,0,0}}p{1cm}<{\centering} | >{\color[RGB]{0,0,0}}p{1.5cm}<{\centering} | >{\color[RGB]{0,0,0}}p{3cm}<{\centering} | >{\color[RGB]{0,0,0}}p{3cm}<{\centering} |  >{\columncolor[RGB]{240,240,240}}>{\color[RGB]{0,0,0}}p{1.5cm}<{\centering} | >{\color[RGB]{0,0,0}}p{3cm}<{\centering}}
\hline\hline
 & \textbf{Data Source} & \textbf{Rational Users} & \textbf{Overpaid Irrational Users} & \textbf{Underpaid Irrational Users} & \textbf{Volatility} & \textbf{Volatility Historical Range}\\ 
\specialrule{0em}{2pt}{1pt}\hline
1 & ICBC & 100\% & 0\%  & 0\% & \textbf{0.1211} &  \multirow{6}{*}{\tabincell{c}{0.037647 - 0.238111  \\(see Table \ref{HistoricalVolatility})}}  \\ 
\specialrule{0em}{2pt}{1pt}\cline{1-6}
2 & ICBC & 70\% & 15\%  & 15\% & \textbf{0.2317}  &     \\ 
\specialrule{0em}{2pt}{1pt}\cline{1-6}
3 & Bitcoin & 100\% & 0\%  & 0\% & \textbf{0.1158}  &      \\ 
\specialrule{0em}{2pt}{1pt}\cline{1-6}
4 & Bitcoin & 70\% & 15\%  & 15\% & \textbf{0.2288}  &      \\ 
\specialrule{0em}{2pt}{1pt}\hline
\end{tabular}
\begin{tablenotes}
\footnotesize
\item Our dataset comprises 400,000 remittance records obtained from the ICBC Macau between January 2022 and December 2022 as well as 400,000 historical market data obtained from Bitcoin between December 2019 and September 2020.
\end{tablenotes}
\end{threeparttable}
}
\end{table}

\section{Conclusion}

Our research begins by examining  the role of blockchain in the financial sector, emphasizing the importance of a Consortium Blockchain for streamlining transactions among institutions. Institutions encounter a significant challenge related to transaction fees impacting block incentives, leading to potential deviant mining behaviors like \textit{Selfish Mining}, \textit{Undercutting}, \textit{Mining Gap}, and \textit{Pool Hopping}. These behaviors are largely influenced by the volatility of block incentives

To tackle these challenges, our study builds upon the DTS strategies proposed by Zhao et al. \cite{9592512}, focusing on reducing block-incentive volatility. We enhance these strategies by integrating the Verkle Tree mechanism, organizing transactions while integrating DTS with current and prospective scalability solutions. This integrated approach alleviates network bandwidth congestion.

Additionally, we abstract the optimization of DTS strategies using the framework of the VRP. By treating DTS as a VRP, we leverage established VRP solution techniques, ranging from heuristic to machine learning algorithms. This application of VRP techniques to the DTS problem allows us to effectively determine near-optimal solutions.

We utilized five optimization algorithms—PSO, DE, GA, CMA-ES, and GBO—to identify the most effective DTS strategy. Extensive experimentation with a historical dataset of 400,000 bitcoin transactions reveals that Verkle-Tree-based DTS strategies can significantly mitigate block-incentive volatility. The most effective strategy identified is the ``Time-based transaction incorporation priority without designated space for small-fee transactions'' (refer to Table \ref{optimization_res}). 

Expanding our methodology and dataset, we assess the DTS strategy's effectiveness using historical market data from Bitcoin and remittance data from the ICBC Macau. These experiments demonstrate the robustness of the proposed DTS strategy, ensuring block incentive stability even amidst irrational user bidding behavior. Consequently, transaction fees can transition from merely incentivizing miners to include transactions in their blocks, to becoming a primary source of block incentives.

\subsection{Future Work}


As discussed in Section 5, the expedited consensus speed of the Fee-based DTS strategy reduces the risk of forking attacks on blocks involving high-fee transactions. This approach can be expanded in the future by integrating additional transaction characteristics like business type, priority, and execution order into the factors affecting block size. This incentivizes miners to prioritize transactions based on specific attributes, thereby reducing the likelihood of block forking. The fee-based DTS strategy could also address Miner Extractable Value (MEV) \cite{daian2020flash} issue which stems from miners' inability to precisely control the order of transactions when included in a block through distributed means.







\end{document}